\documentclass[doublecol,a4paper]{epl2} 
\newcommand{\beq}{\begin{equation}}
\newcommand{\eeq}{\end{equation}}

\usepackage{hyperref}
\usepackage[normalem]{ulem}

\title{Quantum interference in asymmetric superconducting nanowire loops}
\shorttitle{Quantum interference in asymmetric superconducting nanowire loops} %Insert here a short version of the title if it exceeds 70 characters

\author{J. Hudis\inst{1,2}\and J. Cochran\inst{1}\and G. Franco-Rivera\inst{1}\and C.S. Guzman IV\inst{1}\and E. Lochner\inst{2}\and P. Schlottman\inst{2} \and P. Xiong\inst{2} \and I. Chiorescu\inst{1}}
\shortauthor{J. Hudis \etal}

\institute{                    
  \inst{1} Department of Physics and The National High Magnetic Field Laboratory, Florida State University, Tallahassee, Florida 32310, USA\\
  \inst{2} Department of Physics, Florida State University, Tallahassee, Florida 32306, USA
}

\abstract{
Macroscopic phase coherence in superconductors enables quantum interference and phase manipulation at realistic device length scales. Numerous superconducting electronic devices are based on the modulation of the supercurrent in superconducting loops. While the overall behavior of symmetric superconducting loops have been studied, the effects of asymmetries in such devices remain under-explored and poorly understood. Here we report on an experimental and theoretical study of the flux modulation of the persistent current in a doubly-connected asymmetric aluminum nanowire loop. A model considering the length and electronic cross-section asymmetries in the loop provides a quantitative account of the observations. Comparison with experiments give essential parameters such as persistent and critical currents as well as the amount of asymmetry which can provide feedback into the design of superconducting quantum devices.}

\begin{document}

\maketitle

The superconducting order parameter has a well-defined amplitude and phase, and the superconducting states are characterized by long-range phase coherence. The ability to control and manipulate the superconducting phase and quantum interference of superconducting wave functions over long length scales are at the core of numerous functional quantum devices, such as the Superconducting QUantum Interference Device (SQUID)~\cite{clarke2004} and superconducting qubits~\cite{devoret_2004}. An ubiquitous component of such devices are superconducting loops, in which the flux quantization leads to the periodic modulation of a host of observables, for instance its critical temperature or resistance \cite{LittleParksPRL1962}\cite{Vloeberghs_PRL1992} and switching current in the case of SQUIDs \cite{Wernsdorfer2000}\cite{Matveev_PRL2002}\cite{Podd_PRB2007}. Aside from their fundamental interest, interference-based superconducting devices are very sensitive flux detectors~\cite{Cleuziou2006} and are particularly suitable for sensing small amounts of quantum spins~\cite{Yue_APL2017} that have long decoherence times~\cite{bertaina_NatSR2020}, which constitutes an essential aspect for the development of quantum processors.

A hallmark of superconducting loops is the quantum periodicity exhibited by the persistent current with varying external magnetic flux threading the loop. In symmetric devices, the latter has long been evidenced as sinusoidal modulations in junction SQUIDs, corresponding to an out-of-phase sinusoidal pattern for the persistent current and as symmetric triangular modulations in nanowire loops, corresponding to a sawtooth persistent current pattern with a sudden sign-reversal at half-integer quantum fluxes~\cite{clarke2004}. It has long been known that switching current $I_{sw}(\Phi)$ distribution measurements reveal a gradual change between one branch of the sawtooth to the next at these flux values, seen as a smooth change between two $I_{sw}$ modes (see e.g. \cite{Friedrich_SST2019}). In contrast, an asymmetric device was predicted to produce a discontinuous jump in the single-valued switching current $I_{sw}$ at the field of persistent current sign reversal \cite{burlakov_JETP_2014}, although such an effect was not observed.

Asymmetric devices present specific properties. A numerical study based on the time-dependent Ginzburg-Landau theory~\cite{Berdiyorov_PRB2010} revealed that, with increasing injected current at a constant field, an asymmetric device self-regulates in the form of vortex entry, which results in multiple changes of the super-current in contrast to a single jump into the normal state in a symmetric loop. Experimentally, asymmetric flux modulation of $I_{sw}$ has been observed in loop devices with geometric asymmetry \cite{burlakov_JETP_2014} \cite{burlakov_PLA2017} \cite{Sivakov_LTP2014}\cite{Gurtovoi_PLA2020} leading to switching asymmetry \cite{Kanda2007},\cite{Vodolazov2002}; similar behavior was even observed in a geometrically symmetric device, which was attributed to induced thermal inhomogeneity in the device~\cite{Friedrich_SST2019}. Moreover, the switching current relationship to winding number (or loop vorticity) was studied as thermal activated ~\cite{SeguinPRB1992} or quantum \cite{petkovic_NatComm_2016}\cite{Berdiyorov_PRB2010} phase slips and for potential practical applications \cite{Murphy_NJP2017}. In general, a comprehensive measurement and detailed modeling of the multimodal switching currents in asymmetric superconducting loops are necessary for the physical understanding, performance optimization, and proper interpretation of the measurement results of the superconducting quantum interference devices.

We present an experimental investigation and theoretical modeling aimed at elucidating the effects of geometric and electronic asymmetries on the persistent current dynamics in superconducting loops. First, we describe the effect of these two asymmetries (in branch lengths and cross-sections) on the field modulation curve, giving theoretical and experimental evaluation for relevant parameters such as persistent and critical currents, and the location of minima and maxima. In particular, the importance of variations in the electronic (effective) cross-section of a wire is discussed; this is particularly relevant to granular films where device imagery may not directly relate to transport properties. Also, we present the observation of a bi-modal switching behavior which has a field modulation in agreement with our model. The two switching modes are a result of the phase dynamics of the system, for a given set of device parameters and relaxation conditions. Our results are relevant to a wide range of experimental implementations and theoretical interpretation of superconducting quantum interference devices, with or without Josephson junctions \cite{Finkler2010},\cite{Finkler2012}.
 
\begin{figure}%[h]
	\centering
	\onefigure{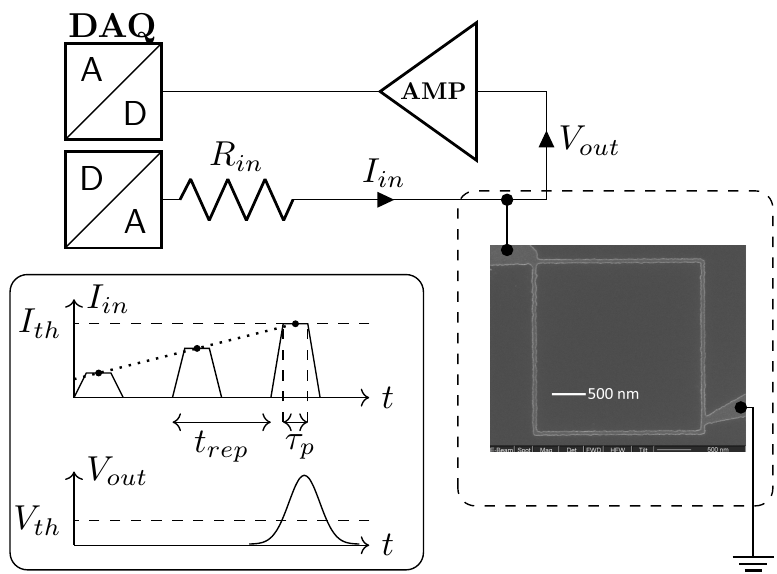}
	\caption{Schematics of the measurement setup. The dashed box shows an SEM image of the aluminium loop, placed in the middle of a superconducting coil at T$\approx$0.3 K. A programmable digital-to-analog converter sends a series of current pulses to the loop (solid box), via an in-line resistor $R_{in}$ and when the detected loop voltage is larger than a set threshold (dashed line), the loop is declared as switched from the superconducting to normal state.} 
	\label{fig1_circ}
\end{figure}

To study the process of phase interference in the case of superconducting waves, first a superconducting loop of size 2.5$\mu$m x 2.5$\mu$m is fabricated on an undoped Si substrate by two-layer (PMGI/PMMA) electron-beam lithography, followed by thermal evaporation of Al and liftoff. Two additional samples were studied to show the reproducible and systematic nature of our theory-experiment agreement, as shown below.  The loop wire has a designed width and thickness of 70~nm and 30~nm respectively. A Scanning Electron Microscopy (SEM) image of a typical sample is shown in Fig.~\ref{fig1_circ} (dashed box). The amount of length asymmetry between the two branches is controlled by the positioning of the second lead, connected to ground. The device presented in Fig.~\ref{fig1_circ} is designed to have only a small amount of geometrical asymmetry. The opposite corner of the square loop is connected to the input-output electronics. The sample is mounted on a holder and inserted in a dilution refrigerator such that the loop is located in the middle of a superconducting coil.

A programmable digital-to-analog voltage source is used to generate a train of current pulses (via an in-line resistor $R_{in}=50$~k$\Omega$) with increasing amplitude, as shown in the solid box of Fig.~\ref{fig1_circ}. The pulses have a length $\tau_p=10\mu$s, repetition time $\tau=100\mu$s and a rise time of about 1.5~$\mu$s. The output voltage is monitored via an amplifier by an analog-to-digital converter (AdWin Gold). The switching of the loop from superconducting to normal state takes place when the loop voltage pulse is larger than a threshold value $V_{th}$ corresponding to a bias current $I_{th}$, as shown in Fig.~\ref{fig1_circ} (here $V_{th}=0.2$~mV). When that happens, the corresponding current and voltage values $I_{in}$ and $V_{out}$ are recorded. After the switch, a device normal state resistance of approximately 140~$\Omega$ is attained, which generates about $\sim\frac{1}{3}\mu$W of heat for a time of the order of $\mu$s $\ll\tau$, before the bias current is reset. The measurement protocol employed ensures that the device is not heated up after a switch.

During an experiment, the magnetic field $\phi$ is fixed and the train of pulses is repeated for $N=1000$ times, in order to obtain a histogram of the switching currents, a method inspired from experiments done with direct-current SQUIDs \cite{Wernsdorfer2000}\cite{Friedrich_SST2019}\cite{SeguinPRB1992}. The flux is then changed by 34.5~m$\Phi_0$ and the procedure is repeated, for the desired range of fields, here scanned from negative to positive values. Since the device is fully switched after each pulse, it is expected that a switching histogram reveals all possible outcomes at each field, without the possibility of a field hysteresis. The temperature of the experiment is set at $T\approx0.3$~K, which is about a fifth of the critical temperature $T_c\approx 1.4$~K \cite{LinPRB2020} (the base temperature inside the superconducting magnet bore is $\approx0.1-0.2$~K).

\begin{figure}%[h]
	\centering
	\onefigure[width=\columnwidth]{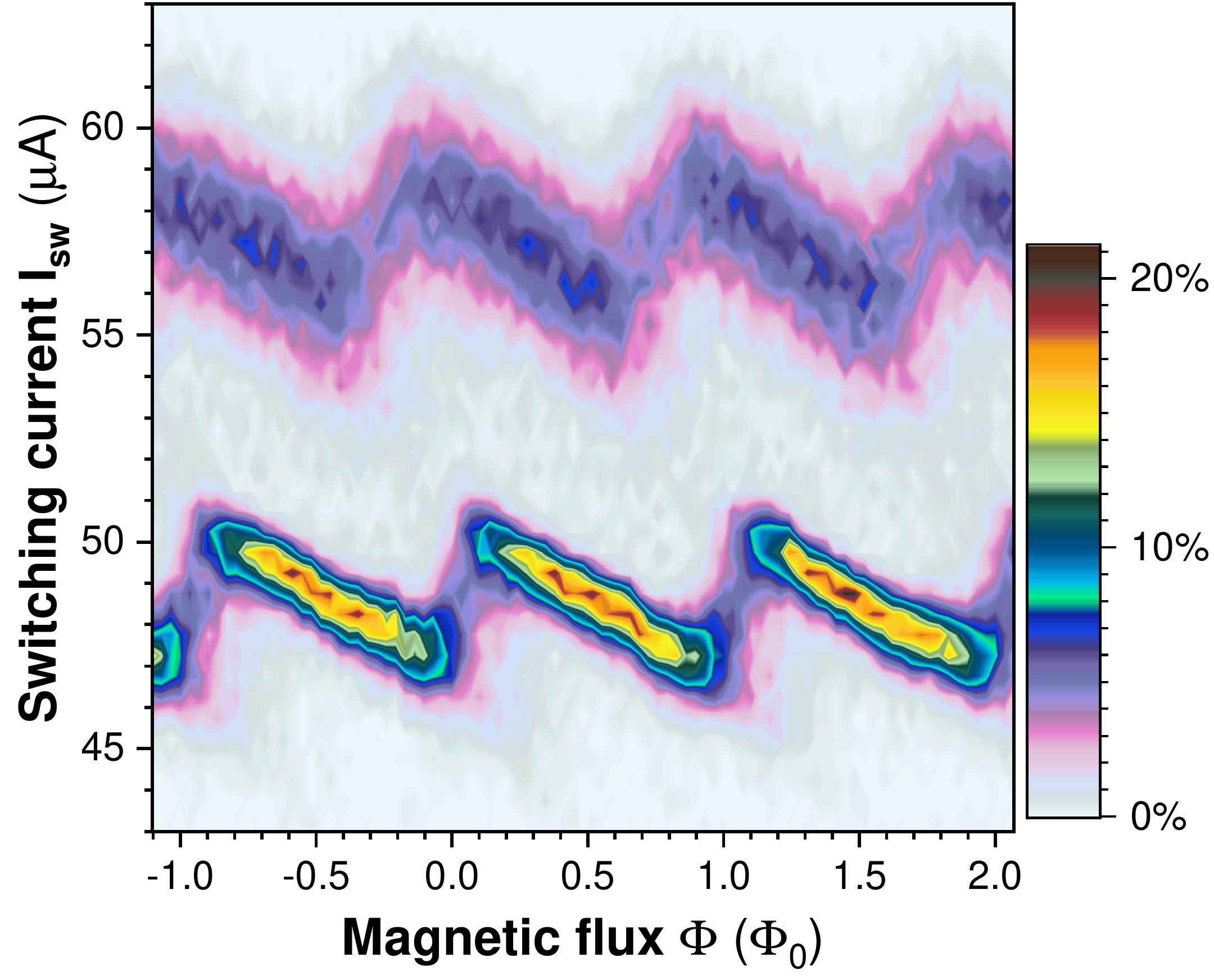}
	\caption{Color map of the switching current histogram as a function of the flux $\phi$ threading the aluminum loop, measured at 0.3~K. The legend shows the counts as percentage of $N$ associated with each color. The modulation curve shows asymmetric positive and negative slopes and the existence of two switching branches.} 
	\label{fig2_contour}
\end{figure}
\begin{figure}%[H]
	\centering
	\onefigure[width=\columnwidth]{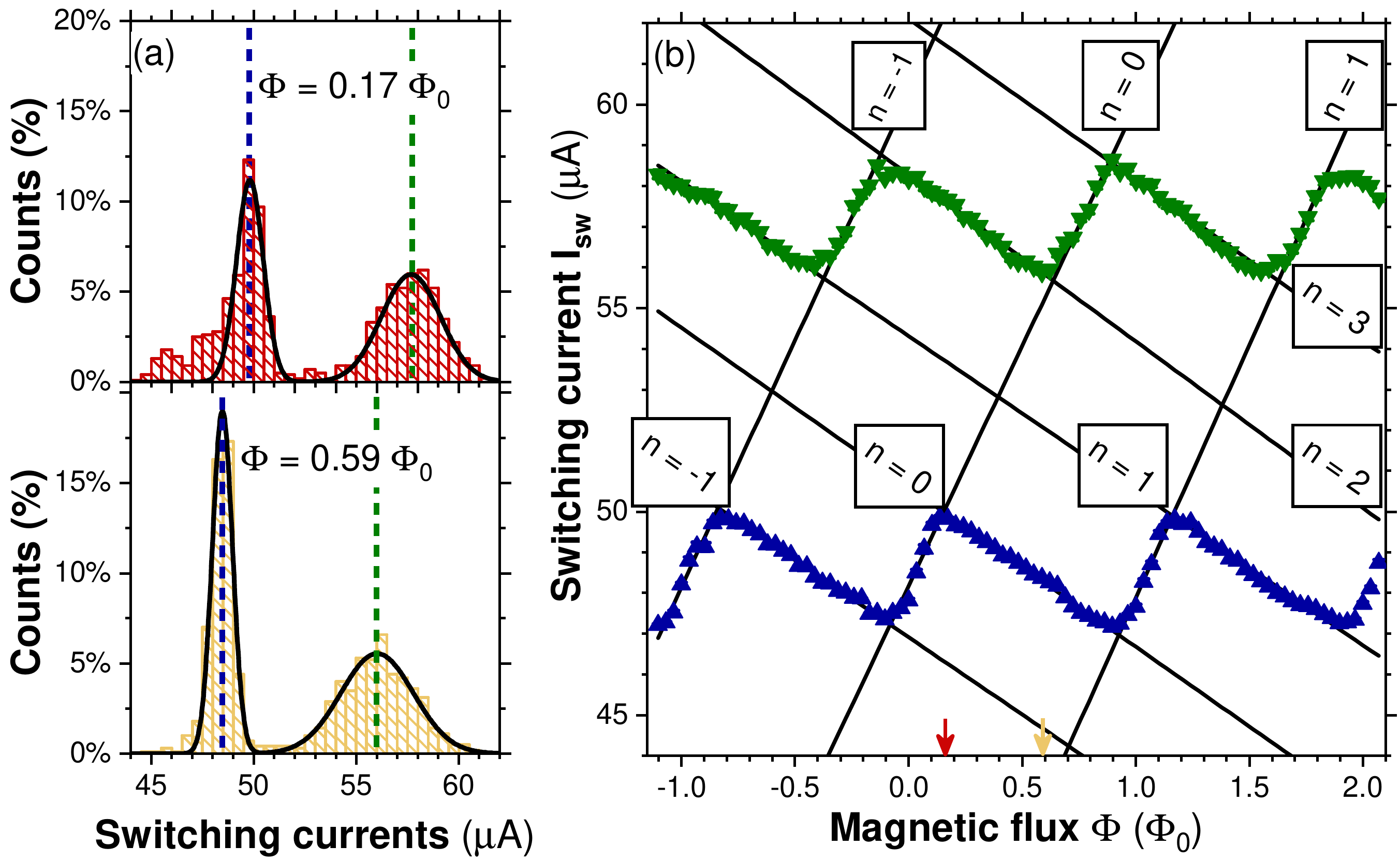}
	\caption{(a) Bi-modal histograms for two values of flux $\phi$, with location shown on the flux axis in panel (b) (dark red and yellow arrows for histograms of same color). The solid line is a double-peaked Gaussian fit that gives the center position and its uncertainty for each mode (dark blue and green dashed lines for the lower and top branch respectively). (b) The fit procedure of panel (a) gives the switching current and uncertainty (with error bars of size similar to dot sizes) for each branch, as a function of $\phi$. The labels indicate the $n$ parameter for the linear fits based on Eq.~\ref{eq_Ic} (dotted lines). } 
	\label{Fig3_model.pdf}
\end{figure}

Using the experimental procedure described previously, series of $N=1000$ values of switching current $I_{sw}$ at fixed fields $\Phi$ are obtained. A waiting time is introduced between consecutive values of the field, to ensure its stability. An $I_{sw}(\Phi)$ series allows to create histograms of the switching current, as is the one presented in Fig.~\ref{fig2_contour}, that represents the modulation curve of a loop threaded by magnetic field. The field period of the modulation is multiplied by the effective area of the loop (the midpoint of the wire is used to approximate the contour) and the result is indeed $\Phi_0$, within the limits of this approximation. The total time to perform such a field scan is $\approx2$~hr during which a small cooling effect is observed on the critical current $I_{c0}$ as a linear drift of $0.6\%$ per $\Phi_0$; this drift is subtracted before obtaining the plot of Fig.~\ref{fig2_contour}. Knowing that $I_{c0}\propto[1-(T/T_c)^2]^{3/2}$ \cite{MydoshPR1965} a cooling of several mK during the experiment is sufficient to explain the observed $I_{c0}$ drift. 

The histograms clearly show two branches for $I_{sw}$, one below and one above $\sim52\mu$A. The lower branch histogram is generally significantly sharper than the top one, but their areas (total count per branch per $\phi$) are very similar, meaning that the device has $\sim50\%$ chance to switch on either of them. Furthermore, a first-neighbor correlation study between consecutive values in a  $I_{sw}$ series indicates that, if the device switched on one branch, the next event is two times more likely to be on the other branch than the same one. An interpretation of such correlation is based on the fact that after a switch, the device may cool off in a metastable minimum of its washboard potential, different from the initial one \cite{SeguinPRB1992}. This leads to the appearance of coreless vortices in the loop (persistent currents) which can shift the $I_{sw}$ modulation curve by one or more periods. The subsequent switching event will therefore have the tendency to be on the other branch. The difference in width between the two modes is in line with the existence of a thermally excited metastable branch; at higher temperatures, the second branch will likely be washed out and the faster relaxation process will reset the system in the initial state.

The bimodal nature of histograms is exemplified in Fig.~\ref{Fig3_model.pdf}(a) for two values of the threaded flux $\phi$ (see also the position of corresponding arrows on the horizontal axis of panel b). In this example, at $\phi=0.17\phi_0$ (dark red) the lower branch has a maximum while at $\phi=0.59\phi_0$ (dark yellow) the top branch has a minimum. The solid black line shows a double-peak Gaussian fit used to extract the location and uncertainty of the modes. The position is indicated with dashed lines (dark blue and green for the lower and top branch respectively).

To study the observed phase interference, we consider two branches that can have different lengths as well as electronic cross-sections. In our devices, the kinetic inductance dominates over the geometrical inductance and any magnetic flux fully penetrates the superconductor. The geometric inductance is estimated using the Fast Henry software; for instance, the device of Fig.~\ref{fig1_circ} has a $L_{geo}=2.1$~pH. The kinetic inductance is given by \cite{vanduzer1981}: $L_{kin}=\mu_0s\lambda^2$ where $s$ is the ratio between loop's length and cross-section, $\lambda^2=\lambda_0^2\xi_0/\ell$ with $\lambda_0=16$~nm the clean penetration depth, $\xi_0=1600$~nm the coherence length \cite{vanduzer1981} and the mean-free path limited by thickness $\ell\approx30$~nm. For the same device, we obtain $L_{kin}\approx95$~pH thus much larger than $L_{geo}$. In this case, the following quantization formula holds: $\oint mv dl=h(n-\frac{\Phi}{\Phi_0})$ where the integral is over the loop contour, $v$ and $m$ are the velocity and mass of electrons, $n$ is an integer, $h$ is Planck's constant, $\Phi_0$ is the flux quantum for Cooper pairs and $\Phi$ is the external flux threading the loop.  The above integral can be split along the two branches of the loop which can be identical or different. First, we follow the length asymmetry model~\cite{burlakov_JETP_2014} to which we add asymmetry in the electronic cross-section (area available for current flow). We note with $L_\pm$ the length of branches where the integral is done along the current direction and opposite to it, respectively. The quantization condition can thus be rewritten as:
\beq
I_{v+}\frac{1+\alpha}{2}-I_{v-}\frac{1-\alpha}{2}=I_p\frac{n\Phi_0-\Phi}{\Phi_0/2}
\label{eq_q}
\eeq
where $\alpha=\frac{L_+-L_-}{L_++L_-}$ is the length asymmetry parameter, $I_{v\pm}$ are the electron velocities in each branch multiplied by a factor $qA_0n_s$, $q=2e$ is the Cooper pair charge, $A_0$ is the average cross-section of the loop, $n_s$ is the Cooper pair density and $I_p=qA_0n_sv$ represents the maximum amount of persistent current that would run in a symmetric loop, situation that occurs when $\Phi$ is a half-integer multiple of $\Phi_0$. The velocity $v$ results from the contour integral above as  $v=\frac{h}{2m(L_++L_-)}$. Note that $I_{v\pm}$ are quantities proportional to carrier velocities and have units of current but in general they are not the actual currents $I_\pm$ running in the two branches. Assuming an asymmetry in the currents cross-sections $A_\pm$ of the two branches, defined by the parameter $\alpha'=\frac{A_+-A_-}{A_++A_-}$ with $A_0=\frac{A_++A_-}{2}$, the currents in the two branches are given by $I_\pm=I_{v\pm}(1\pm\alpha')$. The total current is then:
\beq
I=I_++I_-=I_{v+}(1+\alpha')+I_{v-}(1-\alpha').
\label{eq_I}
\eeq
It is therefore possible to obtain the critical current of each branch, and thus of the device, by solving for $I$ when the carriers approach the critical (depairing) velocity $v_c$. With notations $I_{c0}=2qA_0n_sv_c$ and $\beta=\frac{4I_p}{I_{c0}}$, one solves for $I_{v\mp}$ from Eq.~\ref{eq_q} and take $I_{v\pm}=I_{c0}/2$ in Eq.~\ref{eq_I} to obtain the switching current of each branch:
\beq
\frac{I_{c\pm}}{I_{c0}}=\frac{1-\alpha\alpha'}{1\mp\alpha}\mp\beta\frac{1\mp\alpha'}{1\mp\alpha}\left(n-\frac{\Phi}{\Phi_0}\right).
\label{eq_Ic}
\eeq
The asymmetry parameters $\alpha$ and $\alpha'$ are thus tuning the amount of current flowing in each branch. The quantities $I_{c\pm}$, measured as a function of flux $\Phi$, describe the modulation curve of the switching current, which consists of a series of alternating positive ($I_{c+}$) and negative slopes ($I_{c-}$). The theoretical model allows to calculate several parameters of the loop, such as the slope which determines its sensitivity as flux detector, or the crossing of branches leading to minima and maxima of the modulation curve. For instance, maxima occur at positions $(n-\frac{\alpha}{\beta})\Phi_0$ and currents equal to $I_{c0}$ irrespective of $\alpha'$, while minima occur at $\left[n+\frac{1}{2}-\frac{\alpha}{\beta}+\frac{\alpha-\alpha'}{2(1-\alpha\alpha')}\right]\Phi_0$ and currents $I_{c0}\left(1-\frac{\beta}{2}\frac{1-\alpha'^2}{1-\alpha\alpha'}\right)$.

The measured device (see Fig.~\ref{fig1_circ}) is designed to have a small asymmetry of branches length $\alpha$ and no cross-section asymmetry $\alpha'$. However, the modulation curve of Fig.~\ref{fig2_contour} shows a significant asymmetry between its positive and negative slopes, effect that can be analyzed using our model.

Using Gaussian fits (as described in Fig.~\ref{Fig3_model.pdf}), for each flux $\phi$ one obtains two maxima for $I_{sw}$ as well as their uncertainties which serve as error bars in the representation $I_{sw}(\phi)$. It is thus possible to group points from both branches that belong to the same $n$ parameter in Eq.~\ref{eq_Ic}, as shown with boxed labels in Fig.~\ref{Fig3_model.pdf}. Subsequent linear fits (dotted lines) provide slopes $S^{(n)}_\pm$ and intercepts $I^{(n)}_\pm$ with their uncertainties for each $n$ and type of slope (positive or negative).

In particular, intercepts $I^{(n)}_-$ for $n=-1...3$ and their uncertainties have a linear dependence on $n$, as shown by Eq.~\ref{eq_Ic}, with an intercept $I_{n-}=I_{c0}\frac{1-\alpha\alpha'}{1+\alpha}$. One can thus find experimentally $I_{n-}$ and its uncertainty. Similarly, the values of $I^{(n)}_+$ for $n=-1,0,1$ and their uncertainties, lead to the value of $I_{n+}=I_{c0}\frac{1-\alpha\alpha'}{1-\alpha}$ and its uncertainty. Their values are $I_{n-}=50.57\pm0.07\mu$A and $I_{n+}=48.1\pm0.1\mu$A.  Using the ratio $I_{n-}/I_{n+}$ one can find the value of $\alpha$ and its uncertainty as $\alpha=-0.025$ and $\sigma_\alpha=0.002$ respectively. Hereon, uncertainties on model parameter are found using standard error propagation techniques.

The weighted averages and uncertainties of $S^{(n)}_+$ as well as $S^{(n)}_-$ are noted with $S_+$ and $S_-$ respectively; their values are $S_+=12.09\pm0.07\mu$A/period and $S_-=-3.81\pm0.02\mu$A/period. Their ratio is $\frac{-S_-}{S_+}=\frac{I_{n-}}{I_{n+}}\frac{1+\alpha'}{1-\alpha'}$. Therefore one can find the value of $\alpha'$ and its uncertainty as $\alpha'=-0.538$ and $\sigma_{\alpha'}=0.003$ respectively.

Using either $S_-$ or $S_+$ one can calculate the persistent current (for instance $I_p=\frac{-S_-}{4}\frac{1+\alpha}{1+\alpha'}$) and its uncertainty as $I_p=2.01\mu$A and $\sigma_{Ip}=0.02\mu$A respectively. Similarly, using either $I_{n-}$ or $I_{n+}$ one can calculate the critical current (for instance $I_{c0}=I_{n-}\frac{1+\alpha}{1-\alpha\alpha'}$) and its uncertainty as $I_{c0}=50.0\mu$A and $\sigma_{Ic0}=0.1\mu$A. Using the $I_{c0}$ value, one can estimate the $n_sv_c$ product as $\frac{I_{c0}}{4eA_0}\simeq4\cdot 10^{-32}$~$(m^2s)^{-1}$; we note that the separate evaluations of $n_s$ and $v_c$ depend of factors such as dimensionality, geometry and clean/dirty limit (see \cite{WeiPRB2009,takacs_1973}). The model parameters allow to predict the position of minima and maxima on the modulation curve, using the equations presented above. For instance for the lower branch, one obtains maxima at $(0.156\phi_0, 50\mu\textrm{A})$ and minima at $(-0.084\phi_0,47\mu\textrm{A})$ with a $\phi_0$ periodicity, in very good agreement with the experimental data. We do note that the presence of Earth's magnetic field can give a shift of the zero field (or flux) of $\sim 0.25$~G (or $\sim0.07\Phi_0$) if the chip's normal direction is along north-south; for our setup geometry this could be $\sim0.1$~G which is typically within the uncertainty of $\alpha$. Also, we use the superconducting coil only at small fields, to avoid trapping any extra vortices that could lead to another field shift.   

The model parameters are therefore in the expected range of currents and length asymmetry. For instance, an inspection of sample image in Fig.~\ref{fig1_circ} gives an estimate of $|\alpha|$ of about $0.035$, very close to the measured value. We do note however a striking discrepancy in the designed value of $\alpha'=0$ and measured $\alpha'$, visible as a large asymmetry between the negative and positive slopes of the $I_{sw}$ modulation. This is most likely due to the large aspect ratio of branches leading to an increased probability of having fabrication imperfections along the wires. Surface oxidization and grain boundaries can create an effective narrowing of the current path (electronic cross-section), leading to an increase in $\alpha'$. This type of asymmetry has been previously observed in granular aluminum SQUIDs\cite{Friedrich_SST2019} and the model presented here gives a robust way to quantify it and explain it. 
We note that the discontinuous jumps in the switching current, mentioned in the introductory section and \cite{burlakov_JETP_2014}, were not observed. As expected, the flux modulates the switching currents distribution following Eq.~\ref*{eq_Ic}.
\begin{figure}%[H]
	\centering
	\onefigure[width=\columnwidth]{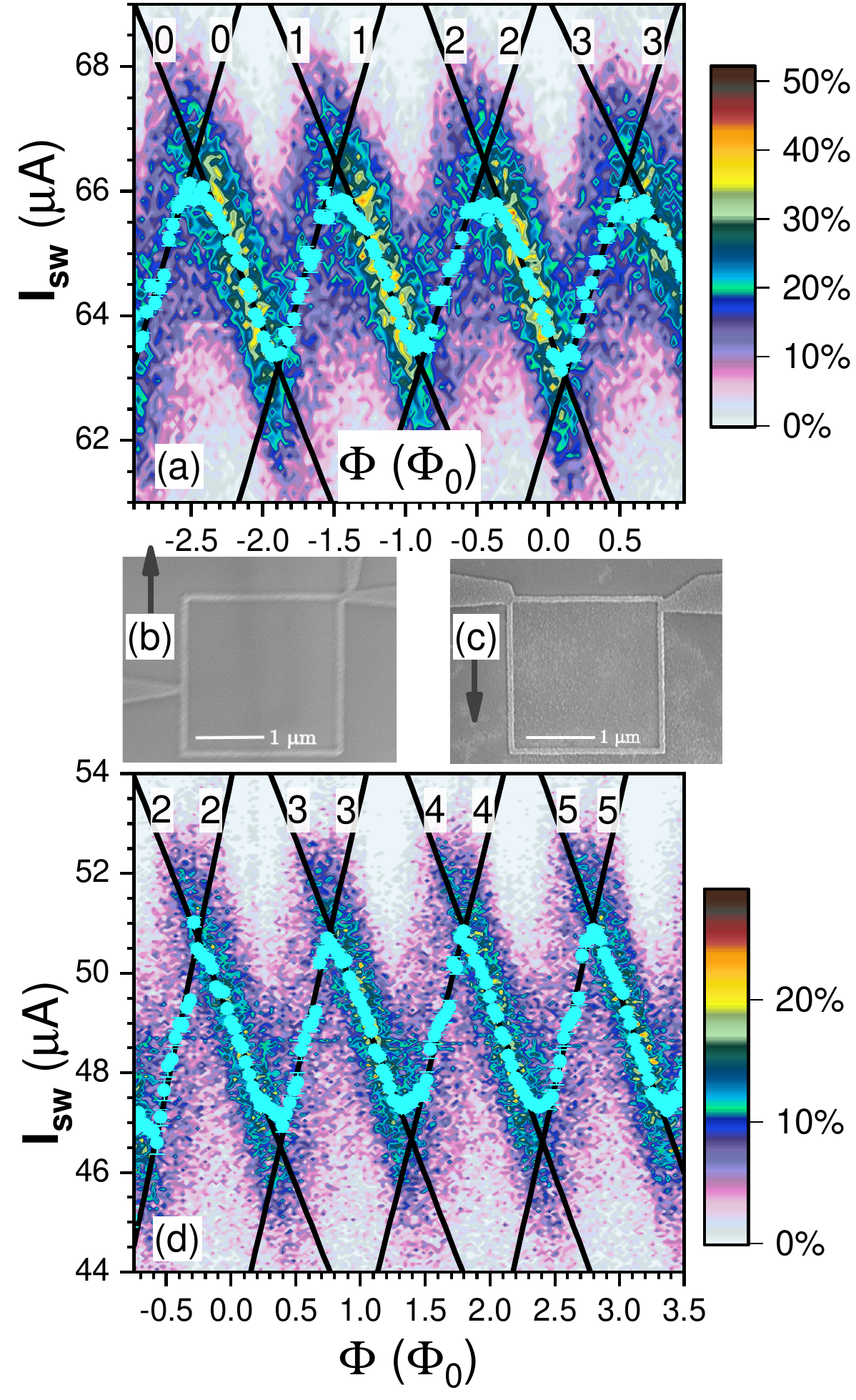}
	\caption{Switching histograms as a function of flux. The red dots represent the location of maxima found using a Gaussian fit (error bars are of similar size as the dots) while the dotted lines represent linear fits using the model Eq.~\ref{eq_Ic} with $n$ labeled by each line. The top plot corresponds to a loop designed to have an asymmetry $\alpha_1=0.21$ while the bottom plot corresponds to a loop with $\alpha_2=0.47$. On the right side, SEM images of the devices are shown.} 
	\label{Fig4_2dev.pdf}
\end{figure}

The effect of $\alpha$ asymmetry is studied experimentally using two devices designated to have medium and high length asymmetry, namely $\alpha_1=0.21$ and $\alpha_2=0.47$ as estimated using the SEM photos shown in Fig.~\ref{Fig4_2dev.pdf}. Using the same experimental and data analysis techniques described above, switching histograms are obtained (see Fig.~\ref{Fig4_2dev.pdf}) together with their corresponding fit parameters $I_\pm^{(n)}$ and $S_\pm^{(n)}$ with $n=0,1,2,3$ and $n=2,3,4,5$ for devices $\alpha_{1,2}$ respectively. The labels $n$ depend strongly on $\alpha$ and $\beta$ since the modulation maxima are located at $(n-\frac{\alpha}{\beta})\phi_0$. The following parameters are obtained: $I_{p1,2}=1.60\pm0.04 / 2.04\pm0.04$~$\mu$A, $I_{c1,2}=66.6\pm0.5 / 50.5\pm0.7$~$\mu$A, $\alpha_{1,2}=0.24\pm0.01 / 0.37\pm0.01$ and $\alpha'_{1,2}=0.1\pm0.02 / 0.16\pm0.02$ for devices $\alpha_{1,2}$ respectively. Once again, we observed a good agreement between the measured $\alpha$ and the one estimated from SEM photos. As it was the case for the first device, an asymmetry of electrical cross-sections $\alpha'$ is observed in these two devices as well, but of a smaller value. Such imperfections are inherent in wires with such long aspect ratio and thus somewhat unpredictable. Nevertheless, it seems plausible to assume that shorter and wider nanowires would have $\alpha'$ values closer to zero. The measurements presented in Fig.~\ref{Fig4_2dev.pdf} are done at $T\sim0.4$~K, higher than that used for the device of Fig.~\ref{fig2_contour} and no second switching branch was observed. This is in line with the view of a metastable state generating the second branch, which relaxes faster to the initial state with the increase of temperature.

Our theoretical and experimental study allows to quantify for a device with two branches the effect of asymmetry on the result of superconducting wave interference. Therefore, it can be used to design and predict essential parameters such as persistent current, modulation depth and flux sensitivity on a given slope, essential to devices such as superconducting qubits and SQUIDs. The model also allows to extract device parameters, like $\alpha$ and $\alpha'$ thus providing feedback on how to improve the relationship between design and fabrication for quantum devices. 

\acknowledgments
Support from the National Science Foundation Cooperative Agreement No. DMR-1644779 and the State of Florida is acknowledged. P.X. acknowledges financial support by NSF grant DMR-1905843.

\bibliographystyle{eplbib}
\bibliography{asym_loop_epl}

\end{document}